\documentclass[10pt]{iopart}
\usepackage[pdftex]{graphicx}
\newcommand{\bc}{\begin{center}}
\newcommand{\ec}{\end{center}}
\begin{document}

\title[Entropic uncertainty relations for electromagnetic beams]{Entropic uncertainty
 relations for electromagnetic beams\footnote{based on the invited talk presented
by one of us (M.A.M.) at the XV Central European Workshop on
Quantum Optics (Belgrade, Serbia, 30 May -- 3 June 2008).}}

\author{Sergio De Nicola,$^1$ Renato Fedele,$^2$
Margarita A. Man'ko$^3$ and Vladimir I. Man'ko$^3$}

\address{${}^1$ Istituto di Cibernetica ``Eduardo Caianiello'' del
CNR Comprensorio ``A. Olivetti'' Fabbr. 70, Via Campi Flegrei, 34,
I-80078 Pozzuoli (NA), Italy\\ ${}^2$ Dipartimento di Scienze
Fisiche, Universit\`{a} ``Federico II'' di Napoli\\ and Istituto
Nazionale di Fisica Nucleare, Sezione di Napoli
\\ Complesso Universitario di M. S. Angelo, Via Cintia, I-80126 Napoli,
Italy\\ ${}^3$ P. N. Lebedev Physical Institute, Leninskii
Prospect 53, Moscow 119991, Russia}

\ead{mmanko@sci.lebedev.ru}
%s.denicola@cib.na.cnr.it~~renato.fedele@na.infn.it~~mmanko@sci.lebedev.ru~~manko@sci.lebedev.ru}

\begin{abstract}
The symplectic tomograms of 2D Hermite--Gauss beams are found and
expressed in terms of the Hermite polynomials squared. It is shown
that measurements of optical-field intensities may be used to
determine the tomograms of electromagnetic-radiation modes.
Furthermore, entropic uncertainty relations associated with these
tomograms are found and applied to establish the compatibility
conditions of the the field profile properties with Hermite--Gauss
beam description. Numerical evaluations for some Hermite--Gauss
modes illustrating the corresponding entropic uncertainty
relations are finally given.
\end{abstract}

%\submitto{\PhysScr}

%\maketitle

\section{Introduction}

The paraxial optical beams propagating in media are known to obey
the Schr\"odinger-like equation \cite{Leon-Fock,Gloge-Marcuse}.
The properties of the paraxial beams described by quantumlike
equations were intensively discussed in both standard optics (see,
for example, \cite{MM1-MM2-MM3,b}) and charged-particle-beam
optics \cite{Renato1-Renato2-Renato3}. In particular, by means of
such quantumlike equations, the paraxial propagation of the
optical-field modes in both free space and in media, where the
refractive index is a given function of the spatial coordinates,
can be described in terms of the Hermite--Gauss or the
Laguerre--Gauss modes \cite{d,e}.

Recently, the tomographic analysis of electromagnetic beams
received a great deal of attention, since it allows for an
accurate characterization of the beam mode profile required when
direct observations are not able to determine the phase of the
optical fields and, therefore, measure of the beam intensity in
all points is, in principle, needed \cite{c}.

In view of a tight connection between the paraxial beam
propagation and quantum mechanics mentioned above, it seems quite
natural to employ a quantum tomographic approach
\cite{Bert-VogRis} to characterize optical beam mode profiles.
Remarkably, to this regard, it was found that the quantum states,
being the solutions to the Schr\"odinger equation, can be mapped
by means of the Radon transform onto the standard probability
distributions called symplectic \cite{ManciniPLA-ManciniFP} and
optical \cite{Bert-VogRis} tomograms. The tomograms were shown
\cite{EPJB06-JMPB-Acta-JRLR-An-TMP-Feynman} to obey the constrains
which correspond to entropic uncertainty relations \cite{Bial-Bur}
for Shannon entropy \cite{Shannon} and R\`{e}nyi entropy
\cite{Renyi} of quantum states.

In this paper, we construct symplectic tomograms of
two-dimensional (2D) optical modes and discuss their properties.
Furthermore, we introduce the concept of tomographic entropy of
such modes and show for them the existence of entropic uncertainty
relations in close analogy to the Heisenberg or
Robertson--Schr\"odinger uncertainty relations. Numerical results
showing the capability of entropic uncertainty relations to
characterize the spatial content of 2D optical modes are
presented.

\section{Symplectic tomography of two-dimensional beams}%NEW
Let us consider a 2D (in Cartesian $x$ and $y$ transverse
coodinates) optical mode field of wavelength $\lambda$ which
satisfies the following paraxial wave equation
\cite{Leon-Fock,Gloge-Marcuse}:
\begin{equation}\label{6}
i\bar\lambda\,\frac{\partial \psi (x,y,z)}{\partial z}=
-\frac{\bar\lambda^2}{2}{\nabla}_\perp^2\psi +
U(x,y,z)\psi(x,y,z),
\end{equation}
where ${\nabla}_\perp^2 = \displaystyle{\frac{\partial^2}{\partial
x^2}} + \displaystyle{\frac{\partial^2}{\partial y^2}}$,
$\bar\lambda = \lambda /2\pi$, $U \propto -\delta n (x,y,z)$ ($n$
being the refractive index close to the $z$ propagation
direction), and $\psi (x,y,z)$ is the complex amplitude of
electromagnetic field. In this equation, the longitudinal
coordinate plays the role of time and $\bar\lambda$ plays the role
of the Planck's constant. The refractive-index profile provides
the possibility to control the modes. In fact, the refractive
index, being dependent on the transversal and longitudinal
coordinates, determines ``the potential energy'' as a function of
the position and time in the corresponding Schr\"odinger-like
equation.

Since we assume a real $U$, the following normalization condition
can be imposed:
\begin{equation}\label{ST1}
\int|\psi(x,y,z)|^2\,dx\,dy=1.
\end{equation}
The tomographic-probability distribution (called also tomogram) is
defined by the Radon transform \cite{Radon1917-Gelfand}
generalized for two dimensions as symplectic tomogram
\cite{DarianoQSO1996} and given in terms of the squared modulus of
the Fresnel integral (\cite{MendesJPA-RitaJRLRGP})
\begin{eqnarray}\label{ST2}
\fl w(X_1,\mu_1,\nu_1,X_2,\mu_2,\nu_2,z)\nonumber\\
\fl =\frac{1}{4\pi^2|\nu_1\nu_2|}
\left|\int\psi(x,y,z)\exp\left[\frac{i}{2}\left(\frac{\mu_1}{\nu_1}x^2
+\frac{\mu_2}{\nu_2}y^2-\frac{2X_1}{\nu_1}x
-\frac{2X_2}{\nu_2}y\right)\right]dx\,dy\right|^2.
\end{eqnarray}
Hereafter, to simplify the notation, we do not show explicitly the
$z$ dependence of the tomogram and related expressions. For
instance, $w(X_1,\mu_1,\nu_1,X_2,\mu_2,\nu_2,z)$ is replaced by
$w(X_1,\mu_1,\nu_1,X_2,\mu_2,\nu_2)$. The tomogram is nonnegative
function of six real variables $X_1$, $\mu_1$, $\nu_1$, $X_2$,
$\mu_2$, and $\nu_2$ and it satisfies the normalization condition
\begin{equation}\label{ST3}
\int w(X_1,\mu_1,\nu_1,X_2,\mu_2,\nu_2)\,dX_1\,dX_2=1.
\end{equation}
Tomogram (\ref{ST2}) can be interpreted as the probability density of
two random variables
%\begin{equation}\label{ST9}
$~X_1=\mu_1x+\nu_1p_x$ and $~X_2=\mu_2y+\nu_2p_y$,
%\end{equation}
where $x$ and $y$ are coordinates of an intersection point in the
transversal plane of the light ray and $p_x$ and $p_y$ are small
angles determining the unit direction vector parallel to the light
ray propagating along the fiber axis.

The tomographic-probability density determines the modulus and
phase factor of the mode function $\psi(x,y)$ due to the inverse
relation
\begin{eqnarray}\label{ST10}
\fl\psi(x,y)\psi^*(x',y')=\frac{1}{4\pi^2}
\int w(X_1,\mu_1,\nu_1,X_2,\mu_2,\nu_2)\delta(\nu_1-x+x')\delta(\nu_2-y+y')\nonumber\\
\fl\times\exp\left[\frac{i}{2}\left(2X_1-\mu_1(x+x_1)+2x_2(y+y')
y^2-\mu_2(y+y')\right)\right]dX_1\,dX_2\,d\mu_1\,d\mu_2\,d\nu_1\,d\nu_2.\nonumber\\
\end{eqnarray}
Tomogram (\ref{ST2}) has the homogeneity property
\begin{equation}\label{ST11}
\fl w(\lambda_1 X_1,\lambda_1 \mu_1,\lambda_1 \nu_1,\lambda_2
X_2,\lambda_2 \mu_2,\lambda_2 \nu_2)
=\frac{1}{|\lambda_1\lambda_2|}
w(X_1,\mu_1,\nu_1,X_2,\mu_2,\nu_2),
\end{equation}
and it can be expressed in terms of optical tomogram depending on
four real variables
\begin{eqnarray}\label{ST12}
\fl w_{\rm
opt}(X_1,\theta_1,X_2,\theta_2)=\frac{1}{4\pi^2|\sin\theta_1\sin\theta_2|}\nonumber\\
\fl\times\left|\int\psi(x,y)\exp\left[\frac{i}{2}\left(\mbox{cot}\,\theta_1(x^2+X_1^2)
+\mbox{cot}\,\theta_2(y^2+X_2^2)-\frac{2ixX_1}{\sin\theta_1}
-\frac{2iyX_2}{\sin\theta_2}\right)\right]dx\,dy\right|^2.\nonumber\\
\end{eqnarray}
In fact, due to the definitions (\ref{ST2}) and (\ref{ST12}), one has
\begin{equation}\label{ST13}
w_{\rm opt}(X_1,\theta_1,X_2,\theta_2)
=w(X_1,\cos\theta_1,\sin\theta_1,X_2,\cos\theta_2,\sin\theta_2)
\end{equation}
and, due to the homogeneity property (\ref{ST11}),
\begin{eqnarray}\label{ST14}
\fl w(X_1,\mu_1,\nu_1,X_2,\mu_2,\nu_2)=\frac{1}{\sqrt{\left(\mu_1^2+\nu_1^2\right)
\left(\mu_2^2+\nu_2^2\right)}}\nonumber\\
\fl\times w_{\rm opt}\left(\frac{X_1}{\sqrt{\left(\mu_1^2+\nu_1^2\right)}}\,,\mbox{arctan}\,
\frac{\nu_1}{\mu_1}\,,\frac{X_2}{\sqrt{\left(\mu_2^2+\nu_2^2\right)}}\,,\mbox{arctan}\,
\frac{\nu_2}{\mu_2}\right).
\end{eqnarray}
Another tomogram called Fresnel tomogram of the light mode in the
optical fiber is given by the integral transform
\cite{fres1-fres2-fres3-fres4}
\begin{eqnarray}\label{ST15}
\fl w_{\rm F}(X_1,\nu_1,
X_2,\nu_2)=\frac{1}{4\pi^2}\frac{1}{|\nu_1\nu_2|}
\left|\int\psi(x,y)\exp\left[
\frac{i(X_1-x)^2}{2\nu_1}+\frac{i(X_2-y)^2}{2\nu_2}\right]dx\,dy\right|^2.\nonumber\\
\end{eqnarray}
Fresnel tomogram $w_{\rm F}$ can be obtained from symplectic
tomogram by putting $\mu_1 = 1$ and $\mu_2 = 1$, i.e.,
%\begin{equation}\label{ST16}
$w_{\rm F}(X_1,\nu_1,X_2,\nu_2)=w(X_1,1,\nu_1,X_2,1,\nu_2)$. %\end{equation}
Furthermore, the following relationship holds:
\begin{equation}\label{ST17}
w(X_1,\mu_1,\nu_1,X_2,\mu_2,\nu_2)=\frac{1}{|\mu_1\mu_2|}w_{\rm F}
\left(\frac{X_1}{\mu_1}\,,\frac{\nu_1}{\mu_1}\,,
\frac{X_2}{\mu_2}\,,\frac{\nu_2}{\mu_2}\right).
\end{equation}
In view of mutual tomogram relations (\ref{ST14}) and
(\ref{ST15}), one can find the mode function $\psi(x,y)$ using
(\ref{ST10}) and express this function either in terms of optical
tomogram or in terms of Fresnel tomogram.

\section{Tomographic entropies of optical beams}

In this section, we consider in detail tomographic entropies
associated with optical beams following the general formalism
developed in \cite{fres1-fres2-fres3-fres4}. There exists the
Shannon construction \cite{Shannon} of entropy associated to the
probability distribution function $P(n)$ of a discrete variable
$n$
\begin{equation}\label{TE1}
H=-\sum_nP(n)\ln P(n),
\end{equation}
which can be easily extended to the classical probability
distribution function of a continuous variable. In particular, we
can use the tomographic-probability distribution (\ref{ST2}) to
introduce the following tomographic entropy for an optical beam:
\begin{eqnarray}\label{TE2} \fl H(\mu_1,\nu_1,\mu_2,\nu_2)=-\int
w(X_1,\mu_1,\nu_1,X_2,\mu_2,\nu_2)\ln
w(X_1,\mu_1,\nu_1,X_2,\mu_2,\nu_2)\,dX_1\,dX_2.\nonumber\\
\end{eqnarray}
The above entropy is a new information characteristic of the
optical-beam profile in media, which yields also the optical
tomographic entropy of the optical beam
\begin{equation}\label{TE3}
\fl H_{\rm opt}(\theta_1,\theta_2)=-\int w_{\rm
opt}(X_1,\theta_1,X_2,\theta_2)\ln w_{\rm
opt}(X_1,\theta_1,X_2,\theta_2)\,dX_1\,dX_2
\end{equation}
and the Fresnel tomographic entropy
\begin{equation}\label{TE4}
\fl H_{\rm F}(\nu_1,\nu_2)=-\int w_{\rm F}(X_1,\nu_1,X_2,\nu_2)\ln
w_{\rm F}(X_1,\nu_1,X_2,\nu_2)\,dX_1\,dX_2.
\end{equation}
All three entropies (\ref{TE2})--(\ref{TE4}) are mutually related.

An interesting property of the tomographic entropies is connected with
the uncertainty relation associated to the light-beam intensities
related to the mode function $\psi(x,y)$ and its Fourier transform
\begin{equation}\label{TE5}
\tilde\psi(p_x,p_y)=(2\pi)^{-1}\int\psi(x,y)\exp^{-i(p_xx+p_yy)}
\,dx\,dy.
\end{equation}
This entropic uncertainty relation reads (see, e.g., \cite{Bial-Bur})
\begin{equation}\label{TE5a}
\fl -\int|\psi(x,y)|^2\ln|\psi(x,y)|^2dx~dy -\int|\tilde\psi(x,
y)|^2\ln|\tilde\psi(x,y)|^2dx~dy\geq 2\ln(\pi e).
\end{equation}
The entropic uncertainty relation was generalized for the
symplectic tomographic entropy as well as for the optical and
Fresnel tomographic entropies \cite{fres1-fres2-fres3-fres4}.

To consider, for example, the optical tomographic entropy, we
introduce the function
\begin{equation}\label{TE6}
\fl R(\theta_1,\theta_2)=H_{\rm opt}(\theta_1,\theta_2)+H_{\rm
opt}(\theta_1+\pi/2,\theta_2+\pi/2)-2\ln(\pi e),
\end{equation}
where $H_{\rm opt}(\theta_1,\theta_2)$ is given by (\ref{TE3}).

According to the new entropic uncertainty relations
\cite{fres1-fres2-fres3-fres4}, the function
$R(\theta_1,\theta_2)$ must be nonnegative for all values of the
angles $\theta_1$ and $\theta_2$, i.e.,
\begin{equation}\label{TE7}
R(\theta_1,\theta_2)\geq 0.\end{equation} This means that, if one
measures the modulus and phase of the mode function $\psi(x,y)$ by
any method, the results of the measurement yield also the function
(\ref{TE6}) which must be nonnegative. Nonnegativity of this
function for all the angles $\theta_1$ and $\theta_2$ can serve as
an extra control of accuracy of the measurements.

\section{Tomography of 2D Hermite--Gauss beams}

We want to illustrate here the tomographic properties of a
two-dimensional Hermite-Gauss (HG) beam of order $(n,m)$ described
by a wave function $\psi_{nm}(x_1,x_2)$. These beams are of a
particular interest since they form a complete set of solutions of
the paraxial equation (\ref{6}) in the case of a linear medium
whose refractive index is a quadratic function of $x_1$ and $x_2$
or in vacuo (note that in (\ref{6}) we have replaced $x$ and $y$
by $x_1$ and $x_2$, respectively). Any other solution can be
written, in principle, as an expansion in terms of HG beams over
all the indices $n$ and $m$. The HG beams play an important role
in the design of spherical resonators since they represent the
resonators' modes having the same wavefronts as a Gaussian beam
but different amplitude distributions.

Let us write the HG beam of the order $(n,m)$ in the following normalized form:
\begin{equation}\label{hg1}
\fl\psi_{nm}(x_1,x_2)=N_{nm}H_n\left(\frac{\sqrt2x_1}{\sigma_0}\right)
H_m\left(\frac{\sqrt2x_2}{\sigma_0}\right)\exp\left[-\frac{x_1^2+x_2^2}{\sigma_0^2}\right],
\end{equation}
where $\sigma_0$ is the width assumed to be the same along the two transverse
directions $x_1$ and $x_2$ and
$N_{nm}=\sqrt{\displaystyle{\frac{2}{\pi\sigma_0^2n!m!2^{n+m}}}}$
is the normalization factor.

The 2D tomogram $w_{nm}$ of HG beam of the order $(n,m)$ is given
by the following six-dimensional expression:
\begin{eqnarray}\label{hg2}
\fl w_{nm}(X_1,X_2,\mu_1,\mu_2,\nu_1,\nu_2)=\frac{1}{(2\pi)^2}\,\frac{1}{|\nu_1\nu_2|}\,
\left|\int\psi_{mn}(\xi_1,\xi_2)\right.\nonumber\\
\fl\times\left.
\exp\left[i\left(\frac{\mu_1}{2\nu_1}\xi_1+\frac{\mu_2}{2\nu_2}\xi_2-\frac{X_1}{\nu_1}\xi_1
-\frac{X_2}{\nu_2}\xi_2\right)\right]d\xi_1\,d\xi_2\right|^2.
\end{eqnarray}
Substituting the HG field given by Eq.~(\ref{hg1}) into Eq.~(\ref{hg2}) and taking into account
the integral relation
\begin{equation}\label{hg3}
\fl\int_{-\infty}^\infty H_n(\alpha y)\exp\left[-\left(y-\beta\right)^2\right]dy=\sqrt\pi
\left(1-\alpha^2\right)^{n/2}H_n\left(\frac{\alpha\beta}{\sqrt{1-\alpha^2}}\right),
\end{equation}
we obtain for tomogram the following expression:
\begin{equation}\label{hg4}
\fl w_{nm}(X_1,X_2,\mu_1,\mu_2,\nu_1,\nu_2)=\frac{1}{(2\pi)^2}\,\frac{1}{|\nu_2\nu_2|}\,
N^2_{nm}\left|I(X_1,\mu_1,\nu_1)\right|^2\left|I(X_2,\mu_2,\nu_2)\right|^2,
\end{equation}
where
\begin{equation}\label{hg5}
\fl
I(X_k,\mu_k,\nu_k)=\sqrt2\left(1-\alpha^2\right)^{n/2}H_n\left(\frac{\alpha_k\beta_k}
{\sqrt{1-\alpha_k^2}}\right)\frac{\exp\left(-X_k^2/4\nu_k^2q_k\right)}{\sqrt{q_k}},
\qquad k=1,2,              \end{equation} and the parameters
$\alpha_k$, $\beta_k$, and $q_k$ are given by
\begin{equation}\label{hg6}
\alpha_k=\frac{\sqrt2}{\sigma_0\sqrt{q_k}}\,,\qquad
\beta_k=\frac{X_i}{2\nu_k\sqrt{q_k}}\,,\qquad
q_k=\frac{1}{\sigma_0^2}-i\,\frac{\mu_k}{2\nu_k}\,.
\end{equation}

In view of the definition given by Eq.~(\ref{hg2}) inserted into
Eq.~(\ref{hg5}) for $k=1,2$, we can write the tomogram
$w_{nm}(X_1,X_2,\mu_1,\mu_2,\nu_1,\nu_2)$ in a form which gives
insight into its physical meaning, namely,
\begin{eqnarray}\label{hg7}
\fl w_{nm}(X_1,X_2,\mu_1,\mu_2,\nu_1,\nu_2)=\frac{1}{|\mu_1\mu_2|}\,
N^2_{nm}\left(\frac{\sigma_0^2}{\sigma_1(z_1)\sigma_2(z_2)}\right)
H_n\left(\sqrt2\frac{X_1}{\mu_1\sigma_1(z_1)}\right)^2\nonumber\\
\fl\times
H_m\left(\sqrt2\frac{X_2}{\mu_2\sigma_2(z_2)}\right)^2
\exp\left[-2\left(\frac{X_1^2}{\mu_1^2\sigma_1(z_1)^2}+\frac{X_2^2}{\mu_2^2\sigma_2(z_2)^2}
\right)\right],
\end{eqnarray}
where the following quantities are defined:
\begin{equation}\label{hg8}
\sigma_k(z)=\sigma_0\sqrt{1+\left(\frac{z_k}{z_0}\right)^2},\qquad
\frac{\mu_k}{\nu_k}=\frac{2\pi}{\lambda z_k}\,,\qquad
z_0=\frac{\pi\sigma_0^2}{\lambda}\,.
\end{equation}

Now we consider a special case $\mu_1/\nu_1=\mu_2/\nu_2=\rho$.

In this case, the tomogram is a function of four phase space
variables only, namely, $X_1$, $X_2$, $\mu$, and $\nu$, and
Eq.~(\ref{hg7}) can be written as follows:
\begin{eqnarray}\label{hg9}
&&w_{nm}(X_1,X_2,\mu,\nu)=\frac{1}{|\mu\nu|}\,
N^2_{nm}\left(\frac{\sigma_0^2}{\sigma^2(z)}\right)
H_n\left(\sqrt2\frac{X_1}{\mu\sigma(z)}\right)^2\nonumber\\
&&\times H_m\left(\sqrt2\frac{X_2}{\mu\sigma(z)}\right)^2
\exp\left[-2\left(\frac{X_1^2}{\mu_1^2\sigma^2(z)}+\frac{X_2^2}{\mu_2^2\sigma^2(z}
\right)\right],
\end{eqnarray}
where, according to Eq.~(\ref{hg8}), we have
$\sigma(z)=\sigma_0\sqrt{1+\left({{z}/{z_0}}\right)^2}$ and
$z={2\pi}/{\lambda\rho}\,,$ with $z_0$ given in Eq.~(\ref{hg8}).

Apart the scaling factor $(|\mu_1\mu_2|)^{-1}$, Eq.~(\ref{hg9})
represents the intensity distribution of the HG beam of the order
$(n,m)$ at a distance $z$ in the transverse plane of rescaled
transverse distances $X_1/\mu$ and $X_2/\mu$. The first part of
Eq.~(\ref{hg9}) gives the well-known dependence of the beam radius
$\sigma(z)$ on the propagation distance $z=2\pi/\lambda\rho$,
which is defined through the second part of Eq.~(\ref{hg9}) in
terms of the ratio $\rho=\mu/\nu$ of the phase space variables. At
$z=0$, the beam radius is given by $\sigma(0)=\sigma_0$.

The wavelength $\lambda$ plays the role of a spatial scaling
parameter in Eqs.~(\ref{hg8}) and (\ref{hg9}). We point out the
formal analogy between the tomographic representation of HG beams
and the paraxial propagation law governing the evolution of the
intensity distribution in the rescaled transverse plane of an
optical HG beam of the order $(n,m)$ and wavelength $\lambda$.
Indeed $z_0$ given by Eq.~(\ref{hg8}) is the well-known confocal
parameter or the depth of the focus of the propagating beam. When
the relation $\mu_1/\nu_1=\mu_2/\nu_2=\rho$ is not satisfied and
the phase space variables assume arbitrary values, the simple
interpretation of tomogram in terms of the intensity distribution
of 2D HG beam at a fixed distance $z=2\pi/\lambda\rho$ does not
more hold. In general, we have two different distances which,
according to Eq.~(\ref{hg8}), are given by
$z_1=2\pi\nu_1/\lambda\mu_1$ and $z_2=2\pi\nu_2/\lambda\mu_2$,
respectively.

However, it is easy to see that Eq.~(\ref{hg8}) is separable along
the two transverse directions $X_1/\mu$ and $X_2/\mu$ and that the
corresponding factors represent the intensity distribution of two
independent 1D HG beams of radii $\sigma_1(z_1)$ and
$\sigma_2(z_2)$ at the distances $z_1$ and $z_2$.

Therefore, the tomogram of 2D HG comprises, in general, the
features of propagation laws corresponding to the projections of
the beam along two orthogonal directions depending on the values
of the phase space variables.

In Figs.~1--3, we present plots of the function
$R(\theta_1,\theta_2)$ calculated for three different cases of 2D
Hermite--Gauss modes. One can see a similar behavior of these
functions for different mode indices and widths. Nevertheless, all
the functions are nonnegative according to the found uncertainty
relations.

\begin{figure}[ht]
\vspace{-4mm}
\bc \includegraphics[width=10cm]{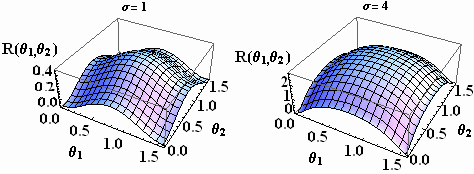} \ec
\vspace{-4mm}
\caption{ %Fig.  1.
Two-dimensional entropic inequality of the Hermite--Gauss beam
H$_{00}$ $(m=0, n=0)$ as a function of the two phase angles
measured in radians with beam width $\sigma= 1$ and 4. }
\end{figure}

\begin{figure}[ht]
\vspace{-4mm}
\bc \includegraphics[width=10cm]{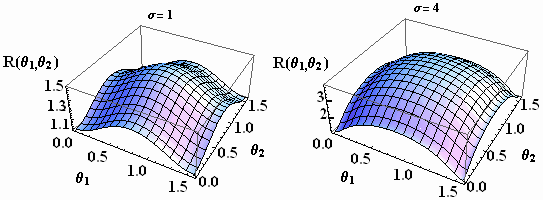} \ec
\vspace{-4mm}
\caption{ %Fig.  2.
Two-dimensional entropic inequality of the Hermite--Gauss beam
H$_{11}$ $(m=1, n=1)$ as a function of the two phase angles
measured in radians with beam width $\sigma= 1$ and 4. }
\end{figure}

\begin{figure}[ht]
\vspace{-4mm}
\bc \includegraphics[width=10cm]{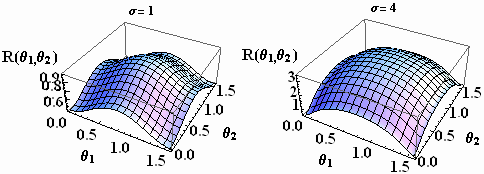} \ec
\vspace{-4mm}
\caption{ %Fig.
Two-dimensional entropic inequality of the Hermite--Gauss beam
H$_{10}$ $(m=1, n=0)$ as a function of the two phase angles
measured in radians with beam width $\sigma= 1$ and 4. }
\end{figure}

\section{Conclusions and discussions}

In this paper, we have shown that the simplectic tomography of
two-dimensional optical beams can be employed to characterize the
spatial distribution of the beam mode along the propagation
direction. With a proper identification of the phase parameters,
the tomographic map reduces to measure the optical intensity
distribution of the optical beam in a given transverse plane.
However, a full tomographic characterization of the beam requires,
in principle, an infinite set of measurements of the spatial
distributions of the light intensity at different points in the
cross section of the optical beam along the propagation direction,
which, in practice, is impossible due to the bandwidth limitations
of the optical detectors. Nevertheless, we have shown that the
tomographic approach can be employed to derive a set of entropic
inequalities in terms of the logarithmic measurement of the beam
intensity. An advantage of this formulation is that it allows one
to catch the relevant features of the propagation beam
characteristics from a tomographic map. Here, the concept of
entropy, well established by statistics and later extended by
Shannon to information theory, plays the role of an integral
descriptor of the tomogram of the optical mode in terms of two
basic optical beam parameters, the width and Rayleigh range.
According to the tomographic-based entropic inequality, a strictly
positive function of two space variables $R(\theta_1,\theta_2)$
can be associated to the optical mode. The numerical results
obtained for the basic Hermite--Gauss modes show clearly that this
entropic-based optical descriptor is sensitive to the beam width
and, in general, tends to increase for higher-order modes. It is
also expected that this formulation can be extended to optical
vortex field where the presence of several defects or vortices is
a manifest signature of high-order optical modes.

\section*{Acknowledgments}

M.A.M. and V.I.M. were partially supported by the Russian
Foundation for Basic Research under Projects Nos.~07-02-00598 and
08-02-90300. M.A.M. thanks the Serbian Academy of Art and Science
for kind hospitality.

\end{document}